\documentclass[journal,10pt]{IEEEtran}

\usepackage{ifpdf}
\usepackage{cite}
\usepackage{amsmath}
\usepackage{array}
\usepackage{stfloats}
\usepackage{cite}
\usepackage{graphicx}
\usepackage{amsmath}
\usepackage{epsfig}
\usepackage{epstopdf}
\usepackage{algorithm}
\usepackage{algorithmicx}
\usepackage{algpseudocode}
\usepackage{blkarray}
\usepackage[table,xcdraw]{xcolor}
\usepackage{float}
\usepackage{amsfonts}
\usepackage{subcaption}
\usepackage{mathrsfs}
\usepackage{amssymb}
\usepackage{graphicx}
\usepackage[utf8x]{inputenc}

\begin{document}
	
\title{Deep-Learning-Aided Detection for Reconfigurable Intelligent Surfaces}
	
\author{Saud Khan,~\IEEEmembership{Student Member,~IEEE}, Komal S Khan,~\IEEEmembership{Member,~IEEE}, Noman Haider,~\IEEEmembership{Member,~IEEE}, and Soo Young Shin,~\IEEEmembership{Senior Member,~IEEE
\thanks{Saud Khan is with the Research School of Electrical, Energy and Materials Engineering, College of Engineering and Computer Science, The Australian National University, Canberra, Australia (Email: saud.khan@ieee.org)}
\thanks{Komal S Khan is with the Wireless Networking Group from the School of Electrical and Information Engineering, The University of Sydney, Sydney, Australia (Email: komal.khan@sydney.edu.au)}
\thanks{Noman Haider is with College of Engineering and Science, Victoria University, Sydney, Australia (Email: noman90@ieee.org)}
\thanks{Soo Young Shin is with the Wireless Emerging and Network Systems Lab, Department of IT Convergence Engineering, Kumoh National Institute of Technology, Republic of Korea (Email: wdragon@kumoh.ac.kr)}% <-this % stops a space
}}
	%\thanks{Manuscript received XXX, XX, 2015; revised XXX, XX, 2015.}}
	
%\markboth{IEEE Journal on Selected Areas in Communications}%
{}
	%{Shell \MakeLowercase{\textit{et al.}}: Bare Demo of IEEEtran.cls for Journals}
	
\maketitle

\begin{abstract}
This paper presents a deep learning (DL) approach for estimating and detecting symbols in signals transmitted through reconfigurable intelligent surfaces (RIS). The proposed network utilizes fully connected layers to estimate channels and phase angles from a reflected signal received through a RIS. Because the proposed network can estimate and detect symbols without any pilot signaling, this method reduces the overhead required for transmission. The improvements achieved by this method are quantified in terms of the bit-error-rate, outperforming traditional detectors. 
\end{abstract}

	
	\IEEEpeerreviewmaketitle
	
\section{Introduction}
\IEEEPARstart{N}{ext-generation} mobile communications networks have begun to be deployed, and compatible handsets will soon be available. The flexibility of wireless communications systems will be improved through the use of unconventional techniques, such as millimeter-wave communications, massive multiple-input multiple-output (MIMO) architectures, and network densification. However, these technologies often cannot be utilized to their full extent because of their high power consumption and poor propagation in harsh environments. These limitations can be mitigated by modifying harsh propagation environments \cite {9086766} \cite{basar2019}. 
	
Researchers have recently proposed controlling the channel medium by converting the environment space into a reconfigurable space known as a reconfigurable intelligent surface (RIS). RIS environments are based on small, low-cost passive elements stacked together in intelligent metasurfaces, which reflect incoming signals towards the receiver with a controllable phase shift. Unlike massive MIMO, RISs are based on a software-defined paradigm, and so they do not require any radio frequency (RF) processing, beamforming, relaying, or other physical RF solutions \cite{nadeem2019intelligent}. 
	
A key feature of RIS-assisted communications making them useful in smart radio environments is their relatively simple implementation, with inexpensive passive elements. However, this feature limits the estimation of channels and phase angles at the receiver end, because RISs cannot perform any RF processing. Therefore, to make RISs as passive as possible to minimize costs, the receiver must estimate the channel and received phase angle with minimal reliance on the RIS \cite{di2019smart}.
	
Deep learning (DL) approaches utilizing deep neural networks (DNNs) have been successfully applied in various fields, such as computer vision and natural language processing. DL has been used for almost every aspect of communications frameworks, in both the physical and network layers, including power allocation \cite{khan2019symbol} and channel estimation \cite{khan2019deep}. Recently, DL was applied to configure the optimal phase angles of RIS meta-elements in indoor communication environments \cite{huang2019indoor}. However, that work assumes that some meta-elements of the RIS are active elements, used for channel estimation, which, reduces the passive nature of RIS. A complexity analysis of the proposed network and its adaptability in a complete channel mismatch scenario also lacks since a perfect channel state information (CSI) was considered only.
	
Most existing studies have not explicitly focused on scenarios with imperfect CSI, or on reducing the complexities of the proposed networks. In recent literature, very few studies have investigated the potential for applying DL to estimation and detection in a RIS-assisted communications system. The discussion below investigates this question and explicitly addresses the complexity of the DNN.
    
The present study makes the following contributions:
\begin{itemize}
        \item We present a novel DL-based detector, called DeepRIS, for use in wireless receivers in RIS-assisted communications scenarios. DeepRIS estimates the channel fading and phase angles from the received signal.
        \item We explain the details of the network's implementation, including the numbers of hidden and activation layers, number of neurons, and learning rate. Additionally, the system-level integration of the proposed network in a RIS-empowered wireless communications scenario is discussed in detail.
        \item  The complexity of the proposed DeepRIS detector is discussed in detail.
        \item Finally, simulation results are presented in terms of the bit-error-rate (BER) in both perfect and imperfect CSI cases, and in scenarios in which the channels and reflecting elements are not matched. The results indicate that DeepRIS achieves a near-optimal BER.
\end{itemize}

\section{Related Work}
\label{relatedwork}
Channel estimation and characterization has been major area of research and innovation in last few decades \cite{van1995channel}. This is because of the understanding that technological advancements can be pushed in the context of hardware and software components of wireless system, and the environment is always considered as an uncontrollable element. The idea of artificially reconfiguring the propagation environment with the help of RIS-enabled communication system is new paradigm shift \cite{basar2019wireless,9086766}. 

The virtualization and softwareization at the core of future wireless systems is now also expanded to modelling wireless environment as an optimization problem. In the context of RIS, Machine Learning (ML) is considered to be one of the key enablers of RIS-enabled wireless systems \cite{zappone2019wireless}. Authors in \cite{di2020smart} have discussed the importance of ML based techniques for RIS-enabled communication networks. Analysis on efficacy of model-based and AI-based and their combine use is presented. Moreover, applications of reinforcement learning for RIS-enabled wireless networks are also presented. For detailed survey of recent works, the work in \cite{di2020smart} extensively presents all the possible case scenarios, important techniques and recent works on RIS-powered smart radio environments. Furthermore, authors in \cite{8052521} demonstrated the efficacy of DL models for channel estimation and signal detection for OFDM based systems. The CSI is estimated via simulated training data to be used later for estimating the received signals.

In \cite{8815428}, a neural-network based technique to control the behavior of elements on RIS is presented. The RIS elements are modelled as nodes and their cross-connections as links in order to control the RIS to elevate the communication performance after the training phase. The ray-tracing simulations were used to assess the performance of the proposed approached. Most of these works consider the active-elements on RIS which can be configured based on software-defined components with an aim to increase the signal strength for a user. However, in this work, we consider passive-elements based RIS which is cost-efficient and instead, propose DeepRIS algorithm to be trained and implemented in the end-device for channel and mismatched phase detection.  

To utilize the RIS effectively and efficiently, some work has been done on the configuration of phase shifts \cite{wu2019beamforming, yu2019miso, han2019large, wu2019intelligent}. A semidefinite relaxation (SDR) method was introduced to optimize the phase shifts of each unit so as to maximize the received signal-to-noise ratio (SNR). Since SDR method is of high computational complexity, a relatively low complexity fix point iteration (FPI) algorithm was proposed in \cite{yu2019miso}. However, when the user is located far away from the BS, the performance loss is relatively high. In \cite{han2019large} and \cite{wu2019intelligent}, the phase shifts of each unit is optimized one by one iteratively in a greedy manner. Thus, it is less efficient for large-scale systems and computationally expensive.

Especially, deep reinforcement learning (DRL) based algorithms have demonstrated their efficiency based on feedback-driven learning and optimization to reconfiguring meta-surfaces \cite{di2019smart}. For instance, authors in \cite{huang2020reconfigurable} proposed DRL based algorithm for transmit beamforming at the source along with the phase shift matrix to increase the sum-rate of multi-user in downlink MISO systems. Another work in \cite {huang2019indoor} employs DL based approach to increase the signal strength for indoor user by training DNN with database of user coordinates and configuring the units of RIS. The primary objective is to maximize the signal strength by directing the signals toward user.    

Authors in \cite{9090876} used DL based channel estimation for mmWave based massive MIMO system. Twin CNNs with nine layers each are fed with received signal to estimate the direct and cascaded channels. However, such approach has higher computational complexity. The summary of key highlights, optimization parameters and considered model for the related literature is given in Table \ref{tab:my-table}. Most of the existing studies have not explicitly considered a trade-off between performance efficiency of proposed methods and computational complexity. In this article, we propose DeepRIS method for symbol detection and channel estimation on receiving device. The DeepRIS has comparatively less computational complexity with better signal detection accuracy. The detailed system model and DeepRIS training and implementation is given in next section.

%%%%%%%%

\begin{table*}[!ht]
\caption{Summary of the important features and implementation details of related works}
\label{tab:my-table}
\resizebox{\textwidth}{!}{%
\begin{tabular}{|l|l|l|l|l|l|}
\hline
\textbf{Related Work, Year}                                                                    & \multicolumn{1}{c|}{\textbf{Key Contributions/Highlights}}                                                                                                                                                                                 & \textbf{For device}                                                        & \textbf{Layer Information}                                                                                                                 & \multicolumn{1}{c|}{\textbf{\begin{tabular}[c]{@{}c@{}}Input,\\ Configuration parameter,\\ Output.\end{tabular}}}                                                           \\ \hline
\begin{tabular}[c]{@{}c@{}}\cite{9090876}\\ 2020\end{tabular}                 & \begin{tabular}[c]{@{}l@{}}DL based channel estimation for RIS \\ base mmWave MIMO Systems (ChannelNet)\\ Estimation of direct and cascaded channels\\ on receiver.\end{tabular}                                                           & Receiving device                                                           & \begin{tabular}[c]{@{}c@{}}Input layer\\ 3 Convolutional layers \\ 2 Fully Connected layers\\ Regression layer\\ Output layer\end{tabular} & \begin{tabular}[c]{@{}l@{}}Received pilot signals,\\ Direct channel and cascading channel,\\ Estimated received signal\end{tabular}                                         \\ \hline
\begin{tabular}[c]{@{}c@{}}\cite{huang2020reconfigurable}\\ 2020\end{tabular} & \begin{tabular}[c]{@{}l@{}}DRL based approach for joint optimization of transmit \\ beamforming and phase angle configurations at RIS.\\ Double DNN to get Q-function from critic and actor networks\\ on input states.\end{tabular} & RIS controller                                                             & \begin{tabular}[c]{@{}c@{}}Input layer\\ 2 Fully Connected layers\\ Output layer\end{tabular}                                              & \begin{tabular}[c]{@{}l@{}}Channel information, Tx/Rx information,\\ Tx power and phase shifts,\\ Q-function.\end{tabular}                                                  \\ \hline
\begin{tabular}[c]{@{}c@{}}\cite{huang2019indoor}\\ 2019\end{tabular}        & \begin{tabular}[c]{@{}l@{}}Trained DNN to maximize received\\ signal strength for indoor user.\\ Ray tracing simulations on 3D indoor\\ environment for performance.\end{tabular}                                                           & RIS elements                                                               & \begin{tabular}[c]{@{}c@{}}Input layer\\ \\ 3 Fully Connected layers\\ Output layer\end{tabular}                                           & \begin{tabular}[c]{@{}l@{}}User location/coordinates,\\ RIS elements configuration for optimal \\ phase angles,\\ Targeting intended user.\end{tabular}                    \\ \hline
\begin{tabular}[c]{@{}c@{}}\cite{8815428}\\ 2019\end{tabular}                 & \begin{tabular}[c]{@{}l@{}}Wireless propagation as NN, different walls\\ as layers, and meta-surfaces as nodes (NNConfig).\\ Feed forward/back-propoagate based \\ Ray tracing simulations for \\ performance validation\end{tabular}       & \begin{tabular}[c]{@{}c@{}}Active elements \\ of met-surfaces\end{tabular} & \begin{tabular}[c]{@{}c@{}}Input layer\\ k - Fully Connected layers\\ Output layer\end{tabular}                                            & \begin{tabular}[c]{@{}l@{}}Impinging signal on met-atom of tile,\\ Tile configuration for retransmission \\ of signal,\\ Received signal power at user device.\end{tabular} \\ \hline
\begin{tabular}[c]{@{}c@{}}\cite{8052521} \\ 2018\end{tabular}                & \begin{tabular}[c]{@{}l@{}}DL based channel estimation for OFDM systems\\ DNN based model trained to minimize the difference\\ between the original symbol and output of neural network\end{tabular}                                       & Receiving device                                                           & \begin{tabular}[c]{@{}c@{}}Input layer\\ 3 Fully Connected layers\\ Output layer\end{tabular}                                              & \begin{tabular}[c]{@{}l@{}}Transmit symbols and channel models,\\ Channel estimation,\\ Received signal.\end{tabular}                                                       \\ \hline
\end{tabular}%
}
\end{table*}

\section{Transmission Through RIS: System Model}
In this section, we provide an overview of the generic model of a RIS-assisted MISO communications system. As shown in Fig. 1, we consider a source ($ S $) equipped with $ M $ antennas to communicate with the RIS comprising $ N $ nearly passive elements, which are reconfigurable and controlled using communication-oriented software. The RIS assists the communication by deliberately reflecting the signal from $ S $ towards the single-antenna destination $ D $ using low-resolution phase shifters, owing to unfavorable propagation conditions. For the ease of implementation, the maximal reflection without power loss at the RIS is considered since the reflecting elements are designed to maximize the reflected desired signal power towards the destination $D$ \cite{wu2019intelligent, huang2019reconfigurable, taha2019enabling}. 
	
\begin{figure}[t]
		\centering
		\includegraphics[width=3.40in, height=2in]{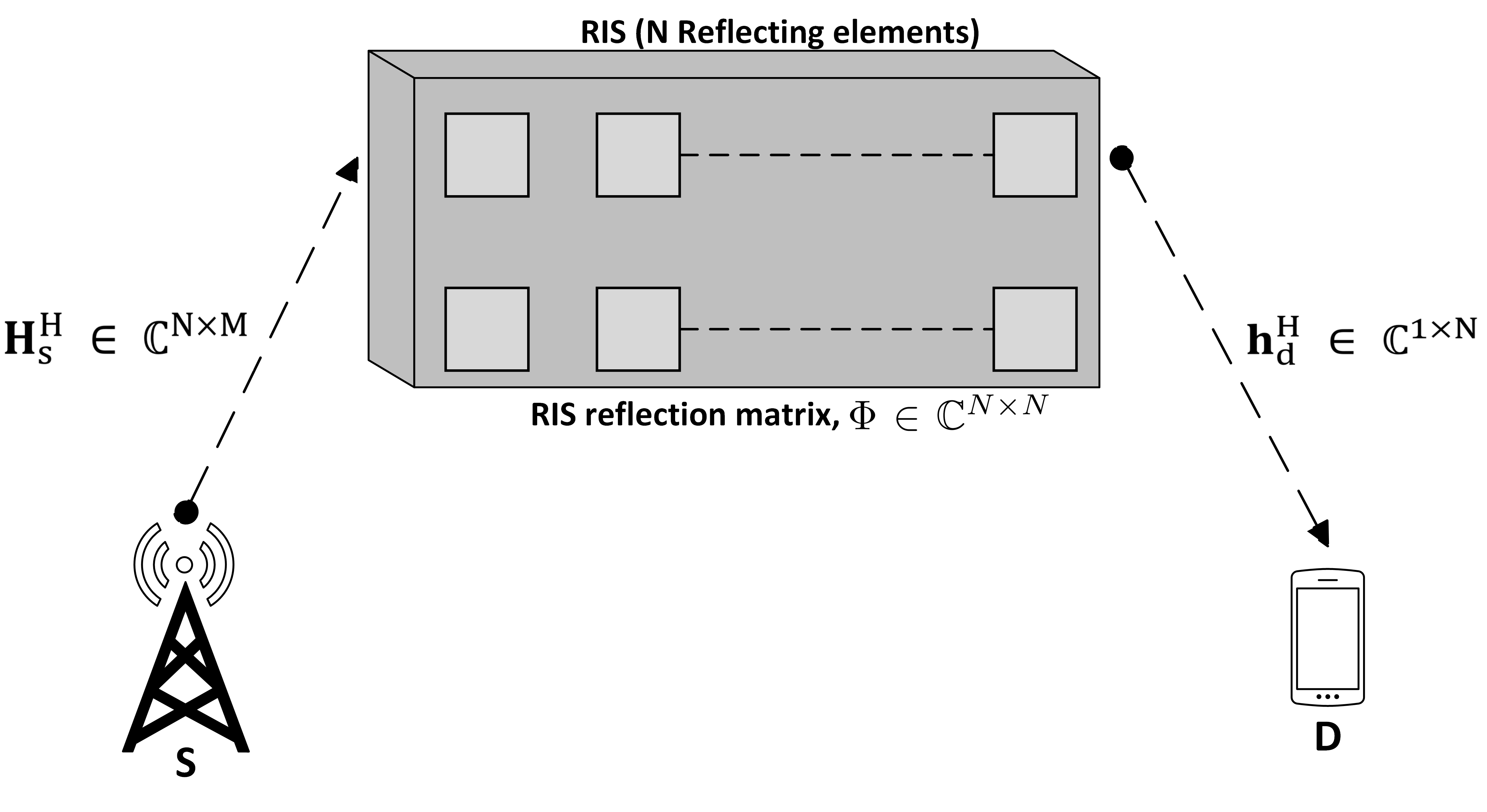}
		\caption{Transmission through an RIS in a dual-hop communication scenario with no line-of-sight path between S and D.}
\end{figure}

Denoting the modulated signal transmitted by $S$ as $\mathbf{x} \in \mathbb{C}^{M \times 1}$, the received downlink signal reflected through the RIS at $D$ can be expressed as,
\begin{equation}
	y_u =	(   \mathbf{h}^H_{d,u} 	\Phi^H 		\mathbf{H}^H_{s,u} )  \sum\limits_{j=1}^U \mathbf{v_j}\mathbf{x_j} + n_u, \; u = 1, 2, ... , U, 
\end{equation}
where $\mathbf{H}^H_{s,u} \in \mathbb{C}^{N \times M}$ denotes the channel matrix from $S$ to RIS, $\mathbf{h}^H_{d,u}  \in \mathbb{C}^{1 \times N}$ denotes the channel vector between the RIS and $D$. $\mathbf{v_j} \in \mathbb{C}^{M \times 1}$ is the beamforming vector at $S$ with the constraint $||\mathbf{v_j}||^2 \leq P_{max}$, where $P_{max}$ is the maximum transmit power of $S$. $\Phi = \text{diag}[\chi_1 e^{j \theta_{1}}, \chi_2 e^{j \theta_{2}}, ... , \chi_i e^{j \theta_{i}}]$ is a diagonal matrix representing the adjustable phase angle induced by the $i_{th}$ reflector of the RIS, where $\chi_i \in [0, 1]$ and $\theta_{i} \in [0, 2\pi]$ represents amplitude reflection factor and the phase shift coefficient on the combined transmitted signal, respectively \cite{nadeem2019intelligent}. $n$ denotes the additive complex Gaussian noise (AWGN) with zero mean and variance $\sigma^2_u$.

From this, the received signal-to-noise ratio (SNR) at $D$ is calculated as
\begin{equation}
	\gamma_u = \frac{| (\mathbf{h}^H_{d,u} 	\Phi^H 		\mathbf{H}^H_{s,u}) \mathbf{v_u} 	|^2} 
	 {\sum_{j \neq u}^U | (\mathbf{h}^H_{d,u} 	\Phi^H 		\mathbf{H}^H_{s,u}) \mathbf{v_j} 	|^2 + \sigma^2_u},
\end{equation}
From (2), it follows that $\gamma$ can be maximized when, 
\begin{equation}
\begin{aligned}
    \text{(P1):}\; \; \; &\underset{\Phi}{\text{max }} \; \; || (\mathbf{h}^H_{d,u} 	\Phi^H 		\mathbf{H}^H_{s,u})	\mathbf{v_u} ||^2\\
            &\text{s.t.} \; \; \; |\chi_i e^{j \theta_{i}}| = 1, \forall \: i = 1, ..., N,
\end{aligned}
\end{equation}
Note that (P1) is an NP-hard problem owing to the non-convexity of the objective function and the unit modulus constraints. An SDR method was proposed in \cite{wu2019intelligent} to solve this problem. However, it is computational expensive with complexity of $\mathcal{O}((N + 1)^6$ \cite{feng2020deep}. From (3), it can be inferred that the destination $D$ will have the maximum received SNR when the phase angles from the reflecting elements of the RIS are known at $D$. Consequently, when the incoming channel phases are known at $D$, the maximum received SNR can be expressed as  
\begin{equation}
	\gamma = \frac{| (\mathbf{h}^H_{d,u} 			\mathbf{H}^H_{s,u}) \mathbf{v_u} 	|^2} {\sigma^2_u} = \frac{A^{2} \mathbf{v_u}} {\sigma^2_u}
\end{equation}
If the phase angles and channel fading are estimated, the received signal can be considered an amplitude gain, represented by $A^2$. Assuming the channel gains as independently Rayleigh distributed random variables, for a large number of reflecting elements $N \gg 1$, from the central limit theorem, $A$ follows a Gaussian distribution with the parameters $E[A] = \frac{N\pi}{4} $ and $VAR[A] = N\left( 1 - \frac{\pi^2}{16}\right)$ \cite{basar2019}. On the other hand, if the phase angles $ \Phi $ and channel fading are unknown at $ D $, the maximum received SNR remains as in (2). From (2) and (4), one can infer that the receiver can only take full advantage of the RIS's reflecting elements if the phase angles are known. If the phase angles are unknown, then the gain is reduced as the number of reflecting elements in the RIS increases.
	
\section{Proposed Deep Learning Detector}
This section details how DeepRIS can be implemented in a RIS-assisted communication scenario. As shown in Fig. 1, the scenario involves two-way channel fading and mismatched phase angles. These challenges are addressed by the DeepRIS DNN, which is trained and then deployed at $ D $ to estimate the received signal without channel estimation. The subsections below detail our analysis of the structure, function, training procedure, and online deployment of the proposed DeepRIS detector. 
	
%\begin{figure}[t]
%	\centering
%	\includegraphics[width=3.40in, height=2in]{figure2}
%	\caption{ Structure of the proposed DeepRIS detector.}
%\end{figure}	

\subsection{Structure of DeepRIS}
DNNs consist of $\ell$ layers having multiple neurons in every layer. The output of the network is a cascade of the nonlinear transformation of input data $\mathbf{I}$ through all the previous layers, which is carried out iteratively. A fully connected layer means that neurons in a layer are connected with the next layer's other neurons. The output of such a network can be expressed as,

\begin{equation}
\mathbf{o} = f (\mathbf{I}, \Theta) = f^{(\ell-1)}	 	(f^{(\ell-2)} 		(\dots f^{(1)}(\mathbf{\mathbf{I}})) )
\end{equation}
where,
\begin{equation*}
f_i = \sigma (w_i \mathbf{I} + b_i)
\end{equation*}
here, each layer at its output has an activation function $\sigma(.)$ and corresponding weights $w_i$ to the input data $\mathbf{I}$ at every training iteration. An additional bias term $b$ is also added to balance the sparse weight balancing during training.  

The proposed DeepRIS network consists of three hidden layers, each of which is a fully connected layer followed by a hyperbolic tangent (tanh) activation function with the range $[-z, z]$. The tanh activation function is preferred over the rectified linear activation function (ReLU) because tanh keeps the negative weights from previous layers intact, whereas ReLU reduces all negative weights to zero. The tanh and ReLU functions can be formally defined as
\begin{equation}
	\begin{split}
	\sigma_{tanh}(z) &= \frac{sinh(z)}{cosh(z)} \\
	&= \frac{e^{z} - e^{-z}}{e^{z} + e^{-z}}\\[10pt]
	\sigma_{ReLU}(z) &= max(0, z),
	\end{split}
\end{equation}
where $z$ denotes the range of the amplitude of the adopted modulation scheme. One can infer that $\sigma_{tanh}$ allows the negative weights to persist, whereas $\sigma_{ReLU}$ reduces all negative weights to zero. This is important to note because the phase angles and the fading channels will also affect the negative modulated constellation symbols. If $\sigma_{ReLU}$ is used as the activation function, then the negative constellation symbols will be reduced to zero, and the network will collapse during training. In contrast, $\sigma_{tanh}(z)$ keeps the negative weights intact, facilitating the training process such that the DNN can accurately estimate the incoming channel and phase angles. 
    
The fully connected layers are designed with the aim that the network can successfully decode the received channel-impaired reflected signals from the RIS with unknown phase angles. This objective is complicated by dual-channel fading from $S$ to the RIS and from the RIS to $D$, as well as unknown phase angles. Therefore, the number of layers and the number of neurons in each layer must be correctly identified. Furthermore, the complexity of the network must be kept at a minimum. Consequently, following a test and trial method, the numbers of neurons in the fully connected layers are set to 500, 250, and 100, in sequence. 
	
\subsection{Model Training}
Before DeepRIS can be deployed as a detector, the model needs to be trained offline with a wide range of instances of channels and phase variances. Random data sequences of bit length $b$ from the source $S$ are generated to simulate transmission vectors. These vectors are then transmitted towards the $N$ reflecting elements of the RIS, and the signal is subject to channel fading $\mathbf{H}^H_{s,u}$. The vectors directed towards the RIS are reflected towards the destination $ D $ with phase angles $ \Phi $, and are again subject to channel fading $\mathbf{h}^H_{d,u}$ along this path. Consequently, the received signal at the destination $ D $ is affected by dual-channel fading and the phase angles of the $ N $ reflector elements of the RIS. 
	
\begin{algorithm}[t]
		\caption{DeepRIS Training Algorithm}
		\textbf{Input:} Original transmitted symbols, channel and phase-affected symbols.\\
		\textbf{Output:} Trained DeepRIS network.
\begin{algorithmic}[1]
			\State Initialization: All parameters including weights, biases, and iterations are set to 0. The validation error threshold is set to $k$.
			\State Produce a set of training patches $\Psi$. 
			\State Process the training patches with simulated channel distortions and phase angles according to (1).
			\While$(k \leq \mathcal{L})$ Train the DNN by minimizing the loss function according to (7).
			\State Update $\mathcal{L}(\Theta)$ according to (8) and (9).
			\EndWhile
			\State \textbf{return:} DeepRIS network
\end{algorithmic}
\end{algorithm}
	
The received channel-faded vectors $y$ and originally transmitted vectors $\mathbf{x}$ are arranged in a training sequence. Because the DeepRIS model's objective is to estimate the originally transmitted symbols at destination $ D $, the model is trained to minimize the difference between the received vector $\mathbf{y}$ and the original transmitted vector $\mathbf{x}$. Therefore, the loss function for the DeepRIS model is defined as
\begin{equation}
	\begin{split}
	\mathcal{L}(\Theta) &=  \frac{1}{N}\sum_{k} \left\|    \mathbf{y(k)} - \mathbf{x(k)}  \right\|^2_2 + \lambda \sum_{k} \Theta_k^2\\
	& =  \frac{1}{N}\sum_{k} \|  \underbrace{ \left[ \mathbf{h}^H_{d, k} 	\Phi^H 		\mathbf{H}^H_{s, k}) \mathbf{v_k}	\right]}_{\xi} 	- 	\mathbf{x(k)} \|^2_2 + \lambda \sum_{k} \Theta_k^2,  
	\end{split}
\end{equation}
where the objective is to estimate $\xi$ over $k$ iterations. This process yields a generic formulation for the estimation of the channel and phase angles in the hidden layers of DeepRIS at the $k_{th}$ iteration. Moreover, $\Theta$ represents the weights and biases of the DeepRIS model, which are updated at every iteration. The parameters $\Theta$ are updated using Adam optimizer, which is defined as
\begin{equation}
	\Theta_{\ell+1} = \Theta_{\ell} - \frac{\eta m_{\ell}}{\sqrt{v_{\ell} + \epsilon}},
\end{equation}
where $\eta$ is the learning rate with which the optimizer defines the step size, and $\epsilon$ is a smoothing term that prevents division by zero. Furthermore, $m_{\ell}$ and $v_{\ell}$ are estimates of the mean and uncentered variance of the gradients, respectively, defined as
\begin{equation}
	\begin{split}
	&m_{\ell} = \delta_{1} m_{\ell-1} + (1-\delta_{1}) \nabla 	\mathcal{L}(\Theta_{\ell})\\[3pt]
	&v_{\ell}  =\delta_{2} v_{\ell-1} + (1-\delta_{2}) \nabla 	[\mathcal{L}(\Theta_{\ell})]^2,
	\end{split}
\end{equation}
where $\delta_{1}$ and $\delta_{2}$ are the decay rates of the moving average. If the gradients in (9) are similar over many iterations, the moving average helps the gradients to gain momentum in a specific direction. Conversely, if the gradients are highly noisy because of sparse training data, then the moving average of the gradient is smaller, and so the parameter updates are also smaller. This relationship is essential in the present case because the DeepRIS detector's inputs will be contaminated by both dual-channel fading and the effects of mismatched phase angles. Therefore, by generating a large and diverse training dataset and training the DNN using Adam optimizer, the DeepRIS detector can reach an optimal solution to the problem. 
	
In order to make the DeepRIS network resilient to overfitting, we include two additional steps during training. Firstly, the $\lambda$ term in (7) is the L2 regularization method, which uses Ridge regression to address the large variances, which are far from the actual value. It adds a squared magnitude of coefficient as penalty term to the loss function during gradient computation and, thus, works efficiently in avoiding the overfitting issue. Second, a dropout layer is incorporated after the final hidden layer of the DeepRIS network, which helps in reducing the overfitting of the network by dropping out random outputs of the layer with a probability $p$. Dropout makes the training process noisy, forcing the nodes within a layer to probabilistically take on, more or less, responsibility for the inputs. This leads to complex co-adaptations, forcing the network layers to co-adapt and correct mistakes from prior layers, making the model more robust \cite{srivastava2014dropout}.
	
\begin{table}[t]
	\caption{Parameters used for simulations}
	\centering
	\begin{tabular}{|c|c|}
		\hline
		\textbf{Parameter}                   	& \textbf{Value}                                 												\\ \hline
		Modulation scheme          			 & 4-QAM                              												 \\ \hline
		Transmit antennas, $M$          	& 32                              														\\ \hline
		Reflecting elements, $N$ 			& 64																				      \\ \hline
		Channel fading               			  & $\mathbf{H}^H_{s,u} , \mathbf{h}^H_{d,u} \sim \mathcal{CN}(0, 1)$               				    \\ \hline
		Pathloss    		   							& $\frac{10^{-2}}{d^{-3.75}}$			  				    \\ \hline	
		AWGN      		   							 & $n \sim \mathcal{CN}(0, SNR)$         			  			         \\ \hline		
		SNR during training        		        & $0\text{dB}$ - $30\text{dB}$                  		            	\\ \hline		
		Batch size        		   					   & 64                 		        											     \\ \hline
		No. of training data samples	  & $7 \times 10^4$ 													     	\\ \hline
		Validation split							 & 	20\%																	     	\\ \hline		
		Validation patience						& 	  50				 													     	\\ \hline		
		Total training steps        			 & 1000                             												   \\ \hline
		Optimizer 							     & Adam, $\delta_{1} = 0.9$, $\delta_{2} = 0.999$         	         \\ \hline
		Regularization 						&  $\lambda = 0.0001$ , $p = 0.5$  \\ \hline
		Learning rate, $\eta$        		   & 0.01                      		        											  \\ \hline		
	\end{tabular}
\end{table}

A training mechanism is implemented for DeepRIS to be adequately trained. First, the network is initialized with empty matrices of weights and biases. Next, random data sequences of size $ \Psi $ are generated and affected using different channel and AWGN intensities and reflected through different $ N $ reflector elements of the RIS. These sequences are subsequently used to train the DeepRIS in an unsupervised manner, where the estimated received signal at the output of the network is updated at every iteration. The concrete learning paradigm is also described in Algorithm 1. By shuffling the channel and AWGN intensities at each iteration, the network maps the effects of the channel and phase angles on the transmitted signal using the nonlinear function approximation in its hidden layers, which help to avoid overfitting of the network.
	
\section{Simulation Results}
Simulation results are used to verify the DeepRIS detector’s performance in terms of the BER. DeepRIS yields an improved performance compared with the conventional least squares (LS) and minimum mean square error (MMSE) estimators. We also discuss the computational complexity of the proposed DeepRIS network during training and deployment, in terms of big-$\mathcal{O}$ notation. Furthermore, we vigorously evaluate the performance of the DeepRIS detector in situations on which it was not trained upon. As such, mismatch scenarios in terms of imperfect CSI, a channel mismatch for small-scale fading, and mismatch $N$ reflecting elements of RIS are modeled, and the performance of DeepRIS is extensively tested. 

For simulations, we consider a composite channel fading based on Rayleigh fading and pathloss component of $3.75$, incorporating large-scale fading. For training, the data samples used are normalized by zero-center using the mean of data. Furthermore, the data samples are shuffled after every epoch during training for generalization. During training, we use 20\% of the training data as a validation test set to the DeepRIS network and set the validation patience to 50 iterations. This essentially means that if the training and validation error is linear for 50 iterations, the network will stop the training process. This is an essential step in the training process as it greatly helps avoid overfitting the network on the training data. 
We use MATLAB for our simulations, and the system parameters used during these simulations are listed in Table II.
	
\begin{figure*}[t]
	\begin{subfigure}{.5\textwidth}
		\centering
		\includegraphics[width=3.45in, height=2.6in]{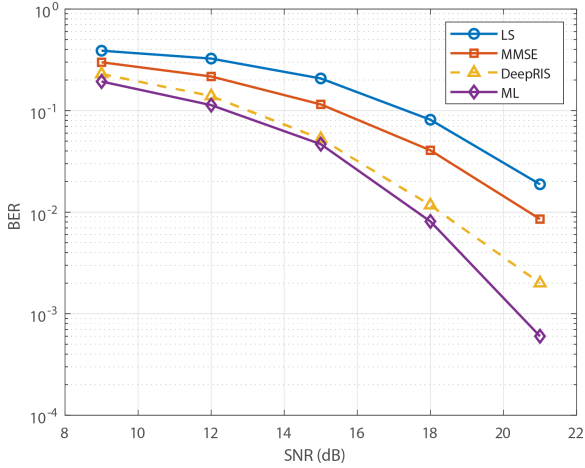}
		\caption{DeepRIS and traditional methods under perfect CSI.}
	\end{subfigure}
	\begin{subfigure}{.5\textwidth}
		\centering
		\includegraphics[width=3.45in, height=2.6in]{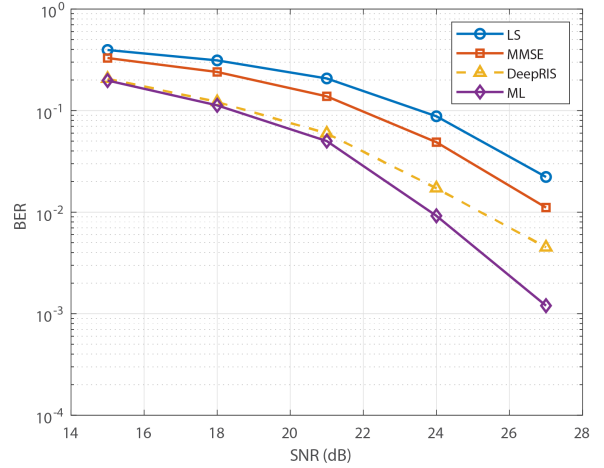}
		\caption{DeepRIS and traditional methods under imperfect CSI.}
	\end{subfigure}		
	\begin{subfigure}{.5\textwidth}
		\centering
		\includegraphics[width=3.45in, height=2.6in]{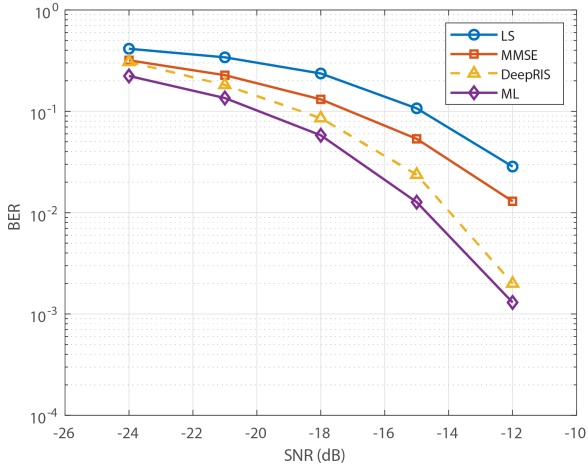}
		\caption{DeepRIS in channel mismatch scenario.}
	\end{subfigure}	
	\begin{subfigure}{.5\textwidth}
		\centering
		\includegraphics[width=3.45in, height=2.6in]{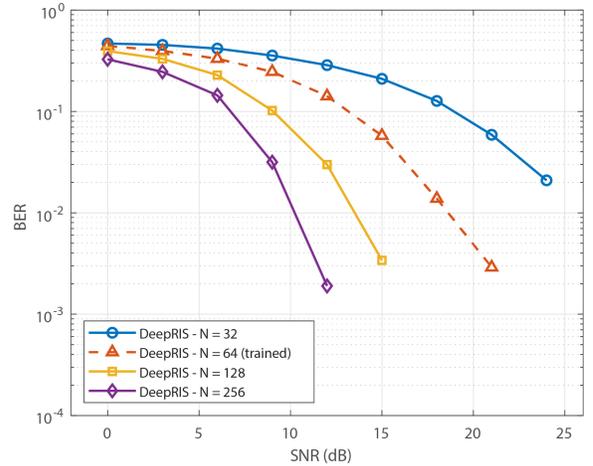}
		\caption{DeepRIS in $N$ reflecting elements mismatch scenario.}
	\end{subfigure}
	\caption{BER Performance of DeepRIS in RIS-assisted communications system.}
\end{figure*}

\subsection{BER Analysis}
We extensively evaluate the BER performance and test various mismatch scenarios to investigate the redundancy of DeepRIS in different situations. Fig. 2 (a) plots the BER curves of DeepRIS and conventional schemes when full CSI is available. As shown, DeepRIS outperforms both state-of-the-art MMSE and LS methods and is comparable to the optimal ML detector. The improved performance of DeepRIS is mainly attributed to the deep architecture of the network, and the extensive training process that helps DeepRIS to regularize to generic variances in channels and phase angles.

In Fig. 2 (b), the BER performance of DeepRIS is plotted when only imperfect CSI is available, and compared with traditional detection schemes. Imperfect CSI means that the receiver is unaware of the channel fading and the receiving phase angles from the RIS. In this case, the transmission is blind. These results show that DeepRIS performs well in situations for which it was not trained. It also outperforms conventional schemes and performs on the level of ML detection. These findings show that DeepRIS is well suited to generic scenarios, and can be deployed without retraining. Moreover, it shows that DL-based detectors can estimate the channel and phase angles of the received signals without explicitly relying on pilot signal-based estimation, which is a crucial issue in RIS-assisted wireless communications.
    
Next, we discuss rigorous performance tests of DeepRIS in situations for which the network was not trained. Although the network is trained using training samples with simulated channel statistics, the network must perform robustly in the case of unforeseen channel fading. Therefore, a channel mismatch scenario was simulated using Nakagami-$m$ fading to analyze how this fading affects the performance of DeepRIS. Nakagami-$m$ fading describes small-scale fading effects and has two parameters that define its shape $m$ and spread $\Omega$. Fig. 2 (c) plots the effects of different channel conditions on DeepRIS in terms of the fading strength $m = 1$ and spreading factor $\Omega = 2$. Furthermore, Fig. 2 (d) analyzes the performance of DeepRIS in the additional untrained scenario in which the number of reflecting elements $N$ of the RIS varies. The results show that the variations in the channel or number of reflecting elements do not significantly degrade the performance of DeepRIS, proving that the network is generalizable.  
	
%Finally, Fig. 3 illustrates the performance of DeepRIS in situations where all the aforementioned challenges are combined, i.e., channel fading is set to small-scale Nakagami-$m$ fading, the number of reflecting elements $N$ of the RIS are mismatched, and imperfect CSI is available. These results show that DeepRIS can adapt to all adversities, and can adapt to all changes occurring during deployment. 
%	
%\begin{figure}[t]
%		\centering
%		\includegraphics[width=3.45in, height=2.6in]{figure3}
%		\caption{BER curves of DeepRIS when combining all adversities.}
%\end{figure}	
	
\begin{figure}[t]
\centering
\includegraphics[width=3.45in, height=2.6in]{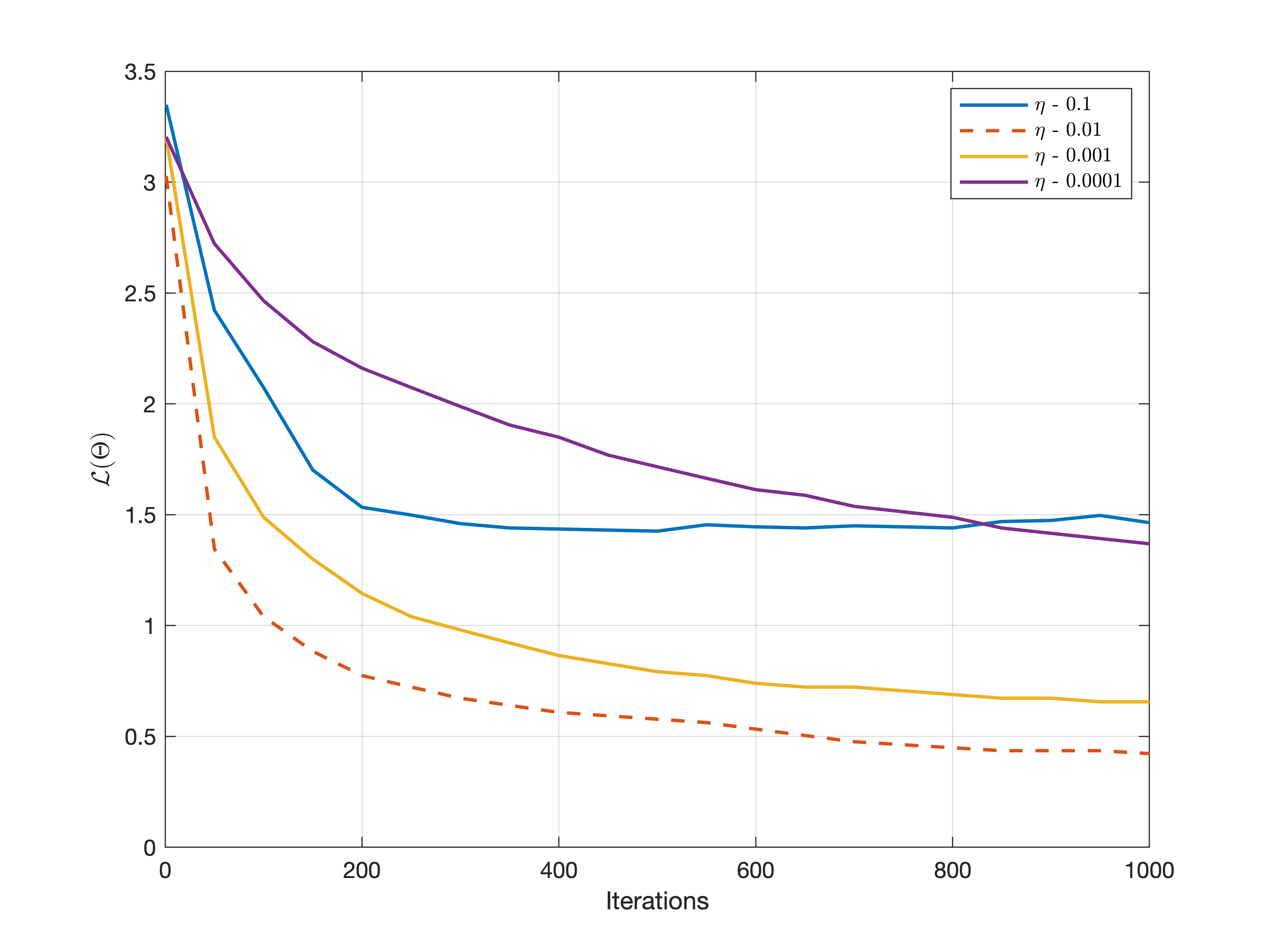}
\caption{Performance analysis of DeepRIS during training for different learning rates.}
\end{figure}	
	
\subsection{Training Analysis}	
Before deployment, the DeepRIS network needs to be trained in a manner that results in less mismatch estimation. In order to achieve this, it is essential to set proper training conditions for the network. In Fig. 3, the variation in training performance of the DeepRIS detector in-terms of different learning rates $\eta$, is shown. It can be inferred that selecting a more significant learning rate during the training backtracks the training performance. This is mainly because of the variational phase angle shifts present in the training data. A more significant learning rate does not allow the network to create a connecting pattern for joint phase and channel estimation. On the other hand, a lower learning rate provides stable training performance, and the pattern is learned accordingly. However, due to the smaller step size, the training time is increased, and the probability of the network getting stuck in a local optima increases as well. As shown in Fig. 3, upon training on different learning rates, a learning rate of $\eta$ of $0.01$ provides a stable training process both in terms of performance and training time. 

Next, the training performance of the DeepRIS detector in-terms of the training and validation error is shown in Fig. 4. As evident, the starting of the training process shows a substantial difference between the training and validation curve. This is because of the random weights assigned to the network at the start of the training. As the training progresses and the network continues to make a pattern, a linear relationship is formed, and a minimal error between the curves is achieved. Consequently, it is shown that the selection of sub-optimal training parameters is essential in utilizing the DeepRIS detector for phase detection and estimation.    

\subsection{Complexity Analysis}
This section analyzes the computational complexity of DeepRIS. The complexity analysis is conducted by calculating the number of operations at each fully connected layer using Big-$\mathcal{O}$ notation. Ignoring the input layer, $p, q, r, \text{and } s$ denote the numbers of nodes in the first, second, third, and fourth layers, respectively. We represent the weights for moving from one layer to another in the form of the matrices $W_{qp}, W_{rq}, \text{and }W_{sr}$, respectively, where $q$, $p$, and so on represent the rows and columns of each weight matrix. The time complexities of operations at every layer are defined as follows:
\begin{itemize}
		\item Time complexity from dense layer 1 $\to$ 2: $\mathcal{O}(q*p)$
		\item Time complexity from activation layer 1 $\to$ 2: $\mathcal{O}(q)$
		\item Time complexity from dense layer 2 $\to$ 3: $\mathcal{O}(r*q)$
		\item Time complexity from activation layer 2 $\to$ 3: $\mathcal{O}(r)$
		\item Time complexity from dense layer 3 $\to$ 4: $\mathcal{O}(s*r)$
		\item Time complexity from activation layer 3 $\to$ 4: $\mathcal{O}(s)$
		\item Time complexity of number of iterations: $\mathcal{O}(k)$
		\item Time complexity of number of training samples: $\mathcal{O}(t)$	
\end{itemize}
Assuming that $t$ training samples pass from layer $1$ to $2$ through activation layer $1$, we see that 
\begin{equation}
\begin{split}
	&\mathcal{O}\left(  q * p * t + q * t \right) \\
	&\mathcal{O}\left(  q *p * (t + 1)\right)\\
	&\mathcal{O}\left(  q * p * t \right)
\end{split}
\end{equation}
Using the same logic for the movements from $q \to r$ and $r \to s$, the time complexity for DeepRIS during training will be
\begin{equation}
	\mathcal{O}\left(  k * t * ( qp + rq + sr ) \right).
\end{equation}
	
The complexity of a network is different during the training and testing stages. During training, a network is trained over many iterations so that it can be generalized and resilient against the overfitting of the data. Hence, the complexity of the training is the network complexity times the number of iterations $k$, as shown in (11). However, during testing and deployment each input to the network is only passed through the network once, with a decision at the output layer. Thus, after training (11) reduces to
\begin{equation}
	\mathcal{O}\left(qp + rq + sr \right),
\end{equation}  
which is the final computational complexity of the deployed DeepRIS network. In comparison with ML detection and MMSE, which have computational complexities of $\mathcal{O}(2^k)$ and $\mathcal{O}(k^3)$, respectively, the proposed DeepRIS detector offers superior performance with lower complexity.

\begin{figure}[t]
\centering
\includegraphics[width=3.45in, height=2.6in]{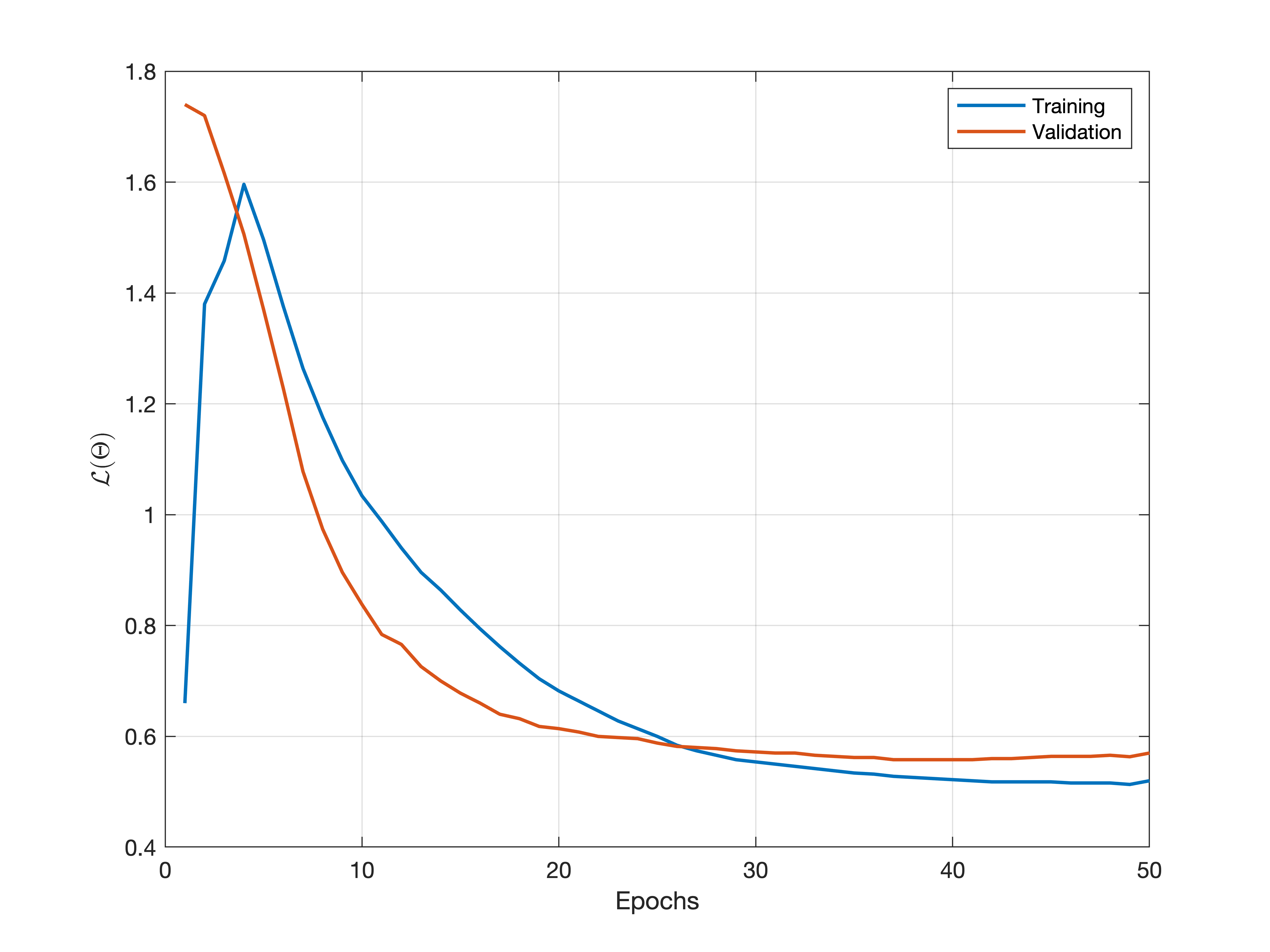}
\caption{Performance of DeepRIS in-terms of training and validation curves.}
\end{figure}

\section{Conclusion}
This work has demonstrated the usefulness of DL methods for estimating channels and phase angles in RIS-assisted wireless communication systems. The DeepRIS detector achieves robust performance in terms of the BER. The model is first trained offline using simulated channel and phase instances. Then, the trained DeepRIS model is deployed to estimate channels and phase angles from the received symbols. The network is easy to train and adaptive to dynamic channel conditions while achieving a lower BER than conventional schemes. 
    
To prepare the model for practical deployment, it must be further generalized to avoid overfitting and perform efficiently in scenarios where channels and phases are mismatched. The preliminary results presented here confirm the potential of DL for estimating channels and phase angles in RIS-assisted wireless communication systems. 
	
	%\section*{Acknowledgment}
	%This work was supported by Priority Research Centers Program through the National Research Foundation of Korea (NRF) funded by the Ministry of Education, Science and Technology (2018R1A6A1A03024003).	
\ifCLASSOPTIONcaptionsoff
\newpage
\fi
	
\bibliographystyle{IEEEtran}
\bibliography{references}
	
\end{document}